\documentclass[epj,referee]{svjour}
\usepackage{graphics}
\begin{document}
\title{Gel-forming patchy colloids and network glass formers: Thermodynamic and Dynamic analogies}
\author{Francesco Sciortino\inst{1} }
%
\offprints{}          
\institute{Dipartimento di Fisica and CNR-INFM-SOFT, Universit\`a di Roma La Sapienza, Piazzale A. Moro 2, I-00185, Rome, Italy}
\date{Received: date / Revised version: date}
%
\abstract{
This article discusses recent attempts  to provide a deeper understanding of the  thermoreversible "gel" state of colloidal matter and to unravel the analogies between gels at the colloidal level and gels at the molecular level, commonly known as network-forming strong liquids.   The connection between gel-forming patchy colloids and strong liquids is provided by the limited valence of the inter-particle interactions, i.e. by the presence of a limit in the number of bonded  nearest neighbors.
} 
\maketitle
\section{Introduction}
Colloidal gels are an arrested state of matter at low density or packing fraction, i.e. an arrested state in which  a negligible fraction of the total volume is occupied by particles. Shear waves of large wavelength can be transmitted trough the gel via the bonded-particle percolating structure.  In the gel, particles are tightly bonded to each other. This requires a thermal energy significantly smaller than the bonding energy, so that the lifetime of the thermo-reversible bonds is comparable or longer than the experimental observation time. 
Despite the apparent simple definition, the essence of the (colloidal) gel state is still under debate, as confirmed by the publication of  several recent reviews, summarizing the work which has been done in the last years\cite{Tra04a,Cip05a,genova,advances,ZaccaJPCM07}.  

In this article I will mostly focus on the expected behavior of 
recently newly synthesized  colloids:  nano and microscopic particles 
with a chemically modulated surface, to provide colloids with valence\cite{Manoh_03,mohovald,Blaad06,Glotz_Solomon}. 
This relevant synthesis effort aims to generate super atoms --- atoms at the nano and micro-scopic level --- 
to reproduce and extent the  atomic and molecular behavior on larger length scale, moving from the 
colloid/hard-sphere-model analogy to more complex cases.
Here I exploit one of these analogies: (thermoreversible) gel-forming patchy colloidal particles and associated liquid, suggesting that network strong glass forming liquids are the atomic and molecular counterpart of  patchy thermoreversible colloidal gels. 

\section{The Role of the Valence}

The connection between  patchy colloidal gels  and network glass-forming liquids is provided by the limited valence of the inter-particle interactions, i.e. by the presence of a limit in the number of bonded  nearest neighbors.  The limited valence can be  brought in or via strong directional interactions (hydrogen bond is the classical molecular example), via interplay of isotropic competing interactions\cite{competing} or due to specific three-body interactions (as in silicon).  At odd  with standard atomic or molecular liquids, in limited valence systems packing is not the driving force determining the structure of the system.
The number of nearest neighbors in the liquid state is significantly smaller than twelve and the hard-sphere model does not play the role of reference system. Crystal structures are characterized by solids with low coordination number, as in the four-coordinated hexagonal ice ($H_20$) and quartz ($SiO_2$) cases. These crystal structures can be considered {\it empty}. The packing fraction of touching hard-spheres arranged in the diamond structure  is indeed only  $\approx 0.34$, less than half of the packing of the hard-sphere face-centered cubic structure.  

\begin{figure}[htbp] 
\resizebox{0.99\columnwidth}{!}{%
 \includegraphics{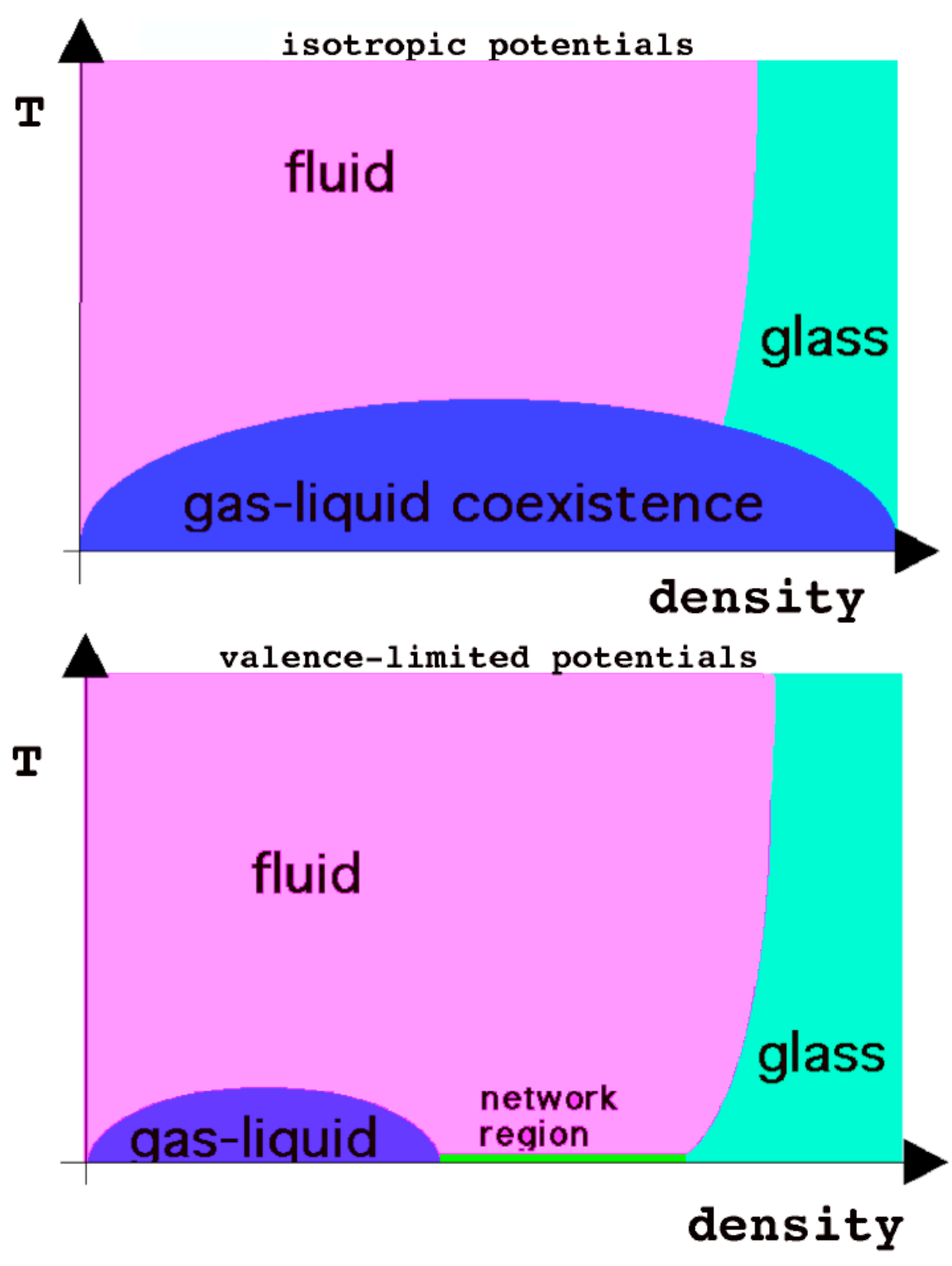}
 }
\caption{Schematic phase diagram  of the disordered phases for the case of particles interacting with non-limited (top) and limited (down) valence potentials. In the standard (top)  case, the glass line hits the gas-liquid spinodal at large densities. In the limited valence case, the shrinking of the gas-liquid unstable region opens up a new region in which a stable extensively bonded network develops. In this region, the dynamics is the one of strong glass-formers. }
   \label{fig:schematic}
\end{figure}

A series of recent studies have focused on the role of the valence in 
determining the field of (meta)stability  of the fluid disordered phase\cite{Zacca1,bian,betheS}.  The field of  (meta)stability can be defined as the region in the temperature-density plane in which the  equilibrium homogeneous fluid (or liquid) phase can exists for a time sufficiently long to allow for an experimental (or numerical) observation. Even if crystallization can be neglected (due to the presence of a nucleation barrier which can sometime be longer than the experimental observation time), the field of stability of the fluid phase is still limited by the glass-line (preventing equilibration of the sample on the experimental time scale)  and by the liquid-gas effective spinodal (preventing the equilibration of the sample in an homogeneous structure).
In the case of particles interacting with spherical symmetric potentials (e.g. simple liquids) or non-symmetric but not valence-limited potentials, the glass line  intersects the gas-liquid spinodal on its right  side (the liquid side)\cite{foffi3}, as schematically shown in Fig.~\ref{fig:schematic}-(top). Thus, at low temperature it is not possible to generated disordered 
{\it homogeneous} (as opposed to phase separated) arrested states. Phase separation covers a large region of the phase diagram.  For square-well type interactions (for which a clear-cut estimate of the packing fraction $\phi$ can be provided since the hard-core distance is properly defined)  the critical packing $\phi_c$ moves from $\phi=0.13$ in the infinite interaction range limit (Kac model) up to $\phi=0.27$ in the infinitesimal interaction range limit (Baxter model)\cite{Miller_03}.   Since the glass line meets the spinodal line at density larger than $\phi_c$, it follows that it is not possible to observe homogeneous arrested states at densities lower than $\phi_c$. 

Reduction of the valence provides an effective mechanism to shrink the gas-liquid coexistence region, opening a large region of intermediate densities in which there is no driving force for phase separation, as schematically shown in Fig.~\ref{fig:schematic}-(bottom)\cite{Zacca1,bian}.  This has been demonstrated  theoretically (via application of the  Wertheim thermodynamic perturbation theory\cite{Werth1,Werth2}) and via simulations for a series of simple patchy models, in which the interaction between particles is composed by an hard-core complemented by several patch-patch interactions, under the condition of single-bond per patch\cite{Zacca1,bian}. It has been shown that the width of the liquid-gas coexistence region progressively reduces on decreasing the valence, disappearing when the average functionality approaches two.  For small valence, the Wertheim theory has been shown to be  very accurate, providing a parameter free description of the system in a wide density and 
temperature region.  The significant shrinking of the gas-liquid region is particularly important  in the present context. Indeed,  it becomes now possible to reach low temperatures without encountering phase separation in conditions such that packing is not an effective arrest mechanism.   This region of low temperature (i.e. of strong long-lived bonding) and of low densities (i.e. in condition where packing is not relevant) can not be accessed in spherical isotropic potentials since it is pre-empted by the phase-separation process.  In the case of limited valence  particles instead, even at intermediate densities, at low $T$ most of the particles are fully bonded and there is no energetic driving force toward a further increase in the local density.  The shrinking of the unstable region depends on the average valence. In the limit of average valence approaching two (which can be realized for example by a binary mixture of particles with two and three patches, in the limit of vanishing concentration of three-functionalized particles) the region of instability approaches zero.For average valence approaching two, a stable homogeneous  {\it empty} liquid phase of vanishing density can be realized.  The structure of the system is composed by long chains of bi-functional particles cross-linked by the three-functional ones.  The possibility of satisfying all possible bonds in a wide region of densities provides thermodynamic stability to the structure.  A detailed discussion of this limiting case can be found in Ref.~\cite{23}.  


Recently it has been reported that universality in the behavior of different limited-valence models can be recovered  using the valence to discriminate between different behaviors.  More precisely, when the leading interaction is provided by bonding (i.e. when the residual interaction is essentially condensed in a hard-core repulsion) the location of the critical point  is controlled by a (valence-dependence) critical value of the second virial coefficient\cite{foffiscaling}.  This scaling generalizes the Noro-Frenkel law of corresponding states for spherical interactions\cite{Noro_00,Vli00a}. The scaling is associated to the possibility of  expanding  the partition function in term of pair-wise bond free energy\cite{foffiPRE}.

The thermodynamic of systems with limited valence is 
characterized by the presence of a temperature at which the
constant volume specific heat $C_V$  has a maximum. The locus of $C_V$ extrema in the phase diagram  is a clear experimentally detectable evidence of the leading role of the bonding interaction in the thermodynamic of the system, since it can be interpreted as arising from the equilibrium between broken and formed bonds.  Moreover,  bonds are very often very well defined interactions. With minor ambiguities, it is possible to  define the presence of a bond between two arbitrary particles (in simulations) or detect the fraction of bonded pairs with spectroscopic measurements.  The study of the connectivity properties of limited valence systems shows the presence of a bond percolation line, passing always slightly above the critical point. At small valence, the $C_V$-max line   lies above the percolation line\cite{due,23}. The opposite behavior is observed at large valence. 

\begin{figure}[htbp] 
\resizebox{0.99\columnwidth}{!}{%
 \includegraphics{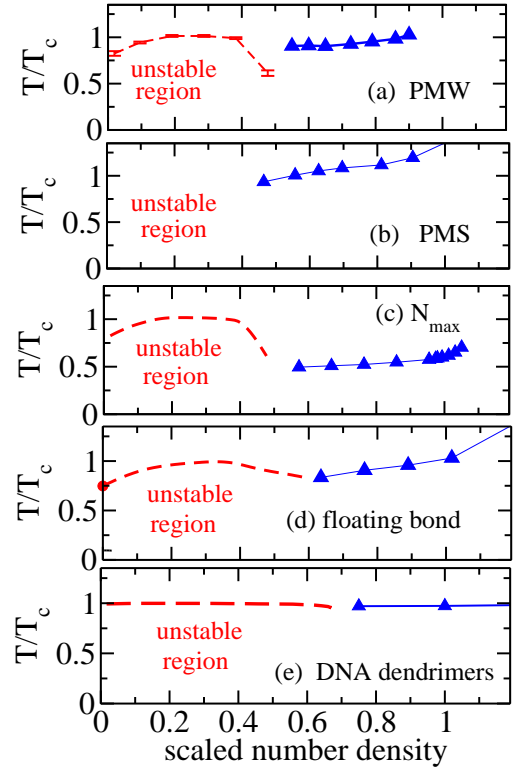}
 }
\caption{Gas-liquid spinodal line and (lowest available)  iso-diffusivity line
   for five different models of patchy particles, in the scaled variables 
   $T/T_c$ and scaled $\rho_{scaled}$.  In all models, the gas-liquid unstable region 
   extends for   $\rho_{scaled}< 0.6$. The iso-diffusivity lines, providing an estimate of  the shape of the glass line, are all mostly parallel to the $T$-axis, bending up only on approaching large densities.   Scaled densities are defined as number of particles 
   divided by the volume, measured in units of the 
   average distance between two bonded particles.    Data are from: (PMW)\cite{simone},  (PMS) \cite{DeMichele_06}, ($N_{max})$\cite{Zacca1,Zacca2},  floating bond model\cite{floatingbond}, the model for four-armed DNA dendrimers\cite{Largo_07}. 
   }
   \label{fig:allpatches}
\end{figure}

I now turn to  discuss properties of simple models with valence four, to make contact with the most common network forming glass formers. 
Several simple patchy models of  four coordinated particles have been recently studied, with the aim of finding out the universal features related to
the valence. These models include the limited valence square-well model ($N_{max}$)\cite{Zacca1,Zacca2}, the primitive model for water (PMW)\cite{Kol_87,simone,romano}, the primitive model for silica (PMS) \cite{Ford_04,DeMichele_06}, the floating bond model\cite{floatingbond}, the model for four-armed DNA dendrimers\cite{dnastarr,Largo_07}.  All these models satisfy the single-bond per patch condition.  Defining the distance between a pair of bonded particles as unit of length and the critical temperature as unit of temperature, all these models are characterized by a similar phase diagram 
(Fig.~\ref{fig:allpatches}).  The gas-liquid unstable region is   limited to scaled densities lower than $\approx 0.6$. Above that density,   the liquid is stable and the structure of the liquid is characterized by  an extended network of four coordinated particles (being the liquid state well within the  percolation region).  In all the single-bond per patch models,  a well defined ground state exists, provided by the geometrically fully bonded configuration.  In all the studied models, the  fully bonded state can be accessed in a region of densities whose width is controlled by the angular constraints\cite{DeMichele_06}.  The approach to the ground state energy along isochoric paths  is well described by an Arrhenius law in temperature. Universal features also control the particle dynamics in these systems. At low $T$, the $T$-dependence of the diffusion coefficient $D$ and of the viscosity $\eta$ approaches an Arrhenius law, the characteristic behavior of so-called strong glass-formers.

An important consequence of the possibility of approaching continuously
the ground state of the system (which in these models is a-priory known, being controlled only by the number of patches per particles and by the 
single-bond per patch condition) provides a proof that  there is no vanishing of the configurational entropy at any finite temperature\cite{Moreno_05,MorenoJCP},  at least in this class of models. More specifically, there is not finite Kauzmann temperature. Even more interestingly, 
in the inherent-structure thermodynamic formalism\cite{Sti88a,JST050515}, the configurational entropy (which counts the number of distinct  configurations with the same energy) can be evaluated  numerically without any approximation\cite{Moreno_05}. Indeed
the number of distinct  configurations  can be formally associated to the number of topological distinct bonding patterns with the same number of bonds and the vibrational entropy to the phase-space explored 
within each bonding pattern. Results suggest that the ground state is degenerate and that the degeneracy is a function of the geometric constraints  limiting the bonding angles. The degeneracy of the ground state reflect the number of distinct fully bonded configurations, differing in the bonded-ring statistics.

\begin{figure}
\resizebox{0.99\columnwidth}{!}{%
  \includegraphics{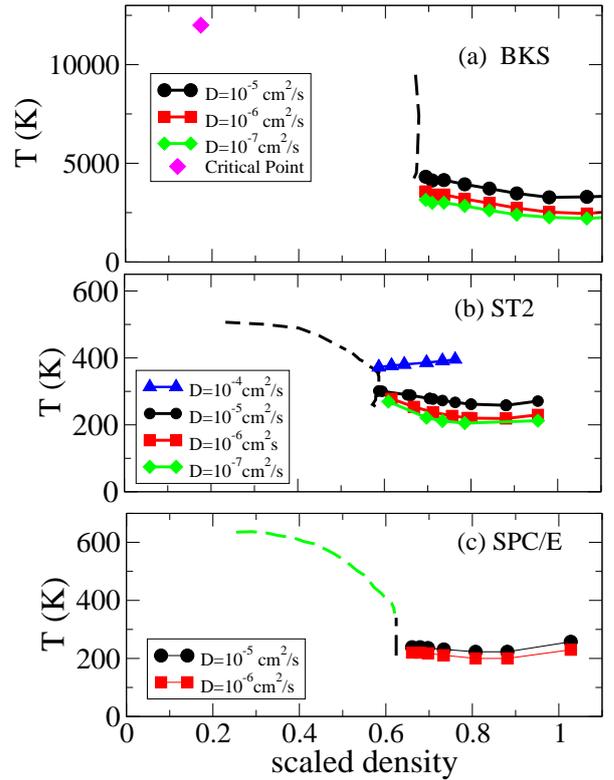}
}
\caption{Phase diagram, including iso-diffusivity lines, of three different models:  (a) the BKS\cite{bks} model for silica. (b)   the ST2\cite{st2} and (c) the SPC/E\cite{spce} models for water.  Dashed lines are the gas-liquid spinodal, full lines are the slowest available iso-diffusivity loci. Different isodiffusivities lines are spaced by one order of magnitude in the value of the diffusion coefficient.  The diagrams are rebuilt using published data. For the BKS potentials data are from Ref.~\cite{saikavoivod,saikavoivod2}, For the ST2 potentials data are from Ref.~\cite{poolejpcm} and for the SPC/E potential
   data are from Ref.~\cite{harrington}.  In the case of water, distances are measured in units of
   2.8 $\AA$  (the HB distance), so that 1.0 g/cm$^3$ correspond to a scaled density of
   $0.734$.  In the case of silica,  distances are measured in units of  ($3.1 \AA$)
   (a characteristic value for the Si-Si distance), so that 1.0 g/cm$^3$ correspond to a scaled density of $0.3$.    }
   \label{fig:bksst2spce}
\end{figure}

The class of valence-four systems includes two of the most abundant liquids on earth, water and silica.  For these materials, several continuous potentials have been developed and extensively studied, providing a reasonable reproduction of several thermodynamic and dynamic properties of the real substances.  These classical potentials, based on electrostatic and Lennard-Jones type of interactions, are able to reproduce the tetrahedral structure characteristic of water and silica under the appropriate density and temperature conditions.  At low $T$, the diffusion coefficient shows an Arrhenius $T$ dependence.  It is thus worth to compare the behavior of these
more complicated and accurate potentials with the generic features observed previously in the simpler one-bond per patch models.  For all models, 
the generic features observed in the patchy colloidal particles models
are confirmed. Even the structure of the tetrahedral network generated by the 
ionic model for silica is astonishingly similar to the one generated by the
short-ranged patchy model for silica\cite{DeMichele_06}.  The gas-liquid unstable region is located in the same region of scaled densities and the iso-diffusivity lines are essentially controlled by temperature (parallel to the $T$-axis),  as shown in Fig.~\ref{fig:bksst2spce}

\section{Conclusions}

The reviewed works point to the relevant role
of the limited valence in the formation of a gel state of matter. 
Dynamic arrest at low densities in an homogeneous system
requires  a very strong inter-particle attraction
(significantly larger than the thermal energy $k_BT$).  Under these conditions, phase separation can be beaten  only when minimization of the energy does not require
an increase in the local density. Interacting patches which can not
form multiple bonds are particularly effective in this respect.
In this spirit, it is possible to rationalize the behavior of
polymeric systems with a small fraction of functionalized monomers, 
the behavior of particles interacting with lock-and-key type of interaction, very often encountered  in biological self-assembly\cite{vecchio,Zhang_04,Workum_06}) as well as  associated liquids and hydrogen bonded compounds\cite{zukosky}.

There are several questions arising from the analogy developed between 
network forming liquids and patchy particles.  In particular I like to call attention on the fact that the reduction of the valence appears to be critical in opening up an intermediate region of densities where bonding  (as opposed to packing) becomes the leading interaction.  The slowing down of the dynamics in this region is found to be Arrhenius and hence typical of the so-called strong glass formers.  Arrest is driven by the formation of a network of long-lived bonds, i.e. by the formation of energetic cages, as opposed to 
the more familiar excluded volume cages  characteristic of fragile glasses.
In the studied models, the slowing down of the dynamics is indeed triggered by the formation of bonds and the Arrhenius dynamics arises when most of the possible bonds are formed. The simplicity of the Arrhenius dynamics 
appears to be connected to the elementary local independent process of bond-breaking. In the simple model investigated\cite{simone,DeMichele_06,dnastarr,Largo_07}, the diffusion process is controlled by the very small concentration of free particles (fully unbonded) diffusing in an empty matrix of fully bonded (and hence not reactive) particles.  In these models the ground state is continuously approached on cooling, proving the absence of any thermodynamic transition associated to the  dynamic arrest. No finite Kauzmann temperature exists. 

In fragile glass-formers, homogeneous arrested states are observed only at very high densities, when excluded volume caging becomes dominant. 
At low density, phase separation prevents the possibility of homogeneous arrest. It is thus tempting to speculate that dynamics can be mostly interpreted in term of hard-sphere dynamics, similarly to the role played by the hard-sphere model as reference system for structural and thermodynamic quantities. In this picture, energy and temperature play a role mainly via the change in the  parameters of the reference state.   If this is the case, fragile and strong liquids are two distinct  (limiting) classes of materials. 

A final comment regards the behavior of liquid water. As discussed, the
gas-liquid unstable region is analogous to the one observed in tetrahedral coordinated patchy models.  Interestingly,  several water models 
are characterized in the region of intermediate densities by an additional
region of instability (the second liquid-liquid critical point\cite{Poo92a}) separating two disordered liquid structures with different densities.  The low-density liquid  has indeed the properties of an {\it empty} liquid structure while the high density one appear to be characterized by a much larger density.
It is not a coincidence, I believe, that dynamics in the  low density liquid  
 is Arrhenius (strong liquid) while it is super-Arrhenius (fragile) in the high density one\cite{PNASstanley}.

I acknowledge support from MIUR-Prin.  I wish to thanks all
colleagues and collaborators (whose names are cited in the References) which have contributed to the results reviewed in this article.


\begin{thebibliography}{10}

\bibitem{Tra04a}
V.~Trappe and P.~Sandk{\"u}hler.
\newblock {\em Curr. Op. Coll. Interf. Sci.}, 8:494--500, 2004.

\bibitem{Cip05a}
L.~{Cipelletti} and L.~{Ramos}.
\newblock {\em J. Phys.:Condens. Matter}, 17:253, 2005.

\bibitem{genova}
F.~Sciortino, S.~Buldyrev, C.~{De Michele}, N.~Ghofraniha, E.~{La Nave},
  A.~Moreno, S.~Mossa, P.~Tartaglia, and E.~Zaccarelli.
\newblock {\em Comp. Phys. Comm.}, 169:166--171, 2005.

\bibitem{advances}
F.~{Sciortino} and P.~{Tartaglia}.
\newblock {\em Advances in Physics}, 54:471--524, 2005.

\bibitem{ZaccaJPCM07}
E.~Zaccarelli.
\newblock {\em J. Phys.: Condens. Matter}, 19:323101, 2007.

\bibitem{Manoh_03}
V.~N. Manoharan, M.~T. Elsesser, and D.~J. Pine.
\newblock {\em Science}, {301}:483--487, 2003.

\bibitem{mohovald}
G.~Zhang, D.~Wang, and H.~M\"ohwald.
  spheres valences.
\newblock {\em Angew. Chem. Int. Ed.}, 44:1--5, 2005.

\bibitem{Blaad06}
A.~van Blaaderen.
\newblock {\em News and Views, Nature}, 439:545--546, 2006.

\bibitem{Glotz_Solomon}
S.~C. Glotzer, M.~J. Solomon, and N.~A. Kotov.
\newblock {\em AIChE Journal}, {\bf 50}:2978, 2004.

\bibitem{competing}
F. Sciortino, S. Mossa, E. Zaccarelli, P. Tartaglia
\newblock {\em Phys. Rev. Lett.} 93:055701, 2004.

\bibitem{Zacca1}
E.~Zaccarelli, S.~V. Buldyrev, E.~La Nave, A.~J. Moreno, Saika-Voivod,
  F.~Sciortino, and P.~Tartaglia.
\newblock {\em Phys. Rev. Lett.}, {\bf 94}:218301, 2005.

\bibitem{bian}
E.~Bianchi, J.~Largo, P.~Tartaglia, E.~Zaccarelli, and F.~Sciortino.
\newblock {\em Phys. Rev. Lett.}, 97:168301--168304, 2006.

\bibitem{betheS}
S.~Sastry, E.~La Nave, and F.~Sciortino.
\newblock {\em J. Stat. Mech.}, 12010, 2006.

\bibitem{foffi3}
G. Foffi, C. De Michele, F. Sciortino and P. Tartaglia.
\newblock {\em Phys. Rev. Lett.}, {\bf 94}, 078301, 2005.

\bibitem{Miller_03}
M.~A. Miller and D.~Frenkel.
\newblock {\em Phys. Rev. Lett.}, {\bf 90}:135702, 2003.

\bibitem{Werth1}
M.S. Wertheim.
\newblock {\em J. Stat. Phys.}, {35}:19--34, 1984.

\bibitem{Werth2}
M.S. Wertheim.
\newblock {\em J. Stat. Phys.}, 35:35, 1984.

\bibitem{23}
E.~Bianchi, P.~Tartaglia, E.~La Nave, and F.~Sciortino.
\newblock {\em J. Phys. Chem. B}, 111:11765--11769, 2007.

\bibitem{foffiscaling}
G.~Foffi and F.~Sciortino.
\newblock {\em J. Phys. Chem. B}, 111:9702--9705, 2007.

\bibitem{Noro_00}
M.~Noro and D.~Frenkel.
\newblock {\em J. Chem. Phys.}, {\bf 113}:2941, 2000.

\bibitem{Vli00a}
G.~A. Vliegenthart and H.~N.~W. Lekkerkerker.
\newblock {\em J. Chem. Phys.}, 112:5364--5369, 2000.

\bibitem{foffiPRE}
G.~{Foffi} and F.~{Sciortino}.
\newblock {\em Phys. Rev. E}, 74(5):050401, 2006.

\bibitem{due}
F.~{Sciortino}, E.~{Bianchi}, J.~F. {Douglas}, and P.~{Tartaglia}.
\newblock {\em J. Chem. Phys.}, 126:4903, May 2007.

\bibitem{Zacca2}
E. Zaccarelli, I. Saika-Voivod, S.~V. Buldyrev, A.~J. Moreno,
  P.Tartaglia, and F. Sciortino.
\newblock {\em J. Chem. Phys.}, {\bf 124}:124908, 2006.

\bibitem{Kol_87}
J.~Kolafa and I.~Nezbeda.
\newblock {\em Mol. Phys.}, {\bf 61}(1):161--175, 1987.

\bibitem{simone}
C. {De~Michele}, S. Gabrielli, P. Tartaglia, and F.
  Sciortino.
\newblock {\em J. Phys. Chem. B}, 110:8064 --8079, 2006.

\bibitem{romano}
F.~{Romano}, P.~{Tartaglia}, and F.~{Sciortino}.
\newblock {\em Journal of Physics Condensed Matter}, 19:F2101, 2007.

\bibitem{Ford_04}
M.~H. Ford, S.~M. Auerbach, and P.~A. Monson.
\newblock {\em J. Chem. Phys.}, {\bf 121}(17):8415--8422, 2004.

\bibitem{floatingbond}
E.~Zaccarelli, F.~Sciortino, and P.~Tartaglia.
\newblock {\em J. Chem. Phys.}, in press (2007).

\bibitem{dnastarr}
F.~W. {Starr} and F.~{Sciortino}.
\newblock {\em J. Phys.: Condens. Matter}, 18:L347--L353, 2006.

\bibitem{Largo_07}
J.~{Largo}, F.~W. {Starr}, and F.~{Sciortino}.
\newblock {\em Langmuir}, 23:5896, 2007.

\bibitem{DeMichele_06}
C.~{De Michele}, P.~{Tartaglia}, and F.~{Sciortino}.
\newblock {\em J. Chem. Phys.}, 125:4710, 2006.

\bibitem{Moreno_05}
A.~J. Moreno, S.~V. Buldyrev, E.~{La Nave}, I.~Saika-Voivod, F.~Sciortino,
  P.~Tartaglia, and E.~Zaccarelli.
\newblock {\em Phys. Rev. Lett.}, {\bf 95}:157802, 2005.

\bibitem{MorenoJCP}
A.~J. Moreno, I.~Saika-Voivod, E.~Zaccarelli, E.~La Nave, S.~V. Buldyrev,
  P.~Tartaglia, and F.~Sciortino.
\newblock {\em J. Chem. Phys.}, {\bf 124}:204509, 2006.

\bibitem{Sti88a}
F.~H. Stillinger.
\newblock {\em J. Chem. Phys.}, 88:7818--7825, 1988.

\bibitem{JST050515}
F.~{Sciortino}.
\newblock {\em Journal of Statistical Mechanics: Theory and Experiment},
  5:15, 2005.

\bibitem{bks}
B.~W.~H. {van Beest}, G.~J. {Kramer}, and R.~A. {van Santen}.
\newblock {\em Phys. Rev. Lett.}, 64:1955--1958, April 1990.

\bibitem{st2}
F.~H. Stillinger and A.~Rahman.
\newblock {\em J. Chem. Phys.}, 60:1545, 1974.

\bibitem{spce}
J.~C. Berendsen, J.~R. Grigera, and T.~P. Straatsma.
\newblock {\em J. Chem. Phys.}, 91:6269, 1987.

\bibitem{saikavoivod}
I.~{Saika-Voivod}, F.~{Sciortino}, and P.~H. {Poole}.
\newblock {\em Phys. Rev. E}, 69(4):041503, 2004.

\bibitem{saikavoivod2}
I.~{Saika-Voivod}, F.~{Sciortino}, T.~{Grande}, and P.~H. {Poole}.
\newblock {\em Phys. Rev. E}, 70(6):061507, 2004.

\bibitem{poolejpcm}
P.~H. {Poole}, I.~{Saika-Voivod}, and F.~{Sciortino}.
\newblock {\em Journal of Physics: Condensed Matter}, 17:L431--L437, 
  2005.

\bibitem{harrington}
S.~Harrington, P.H. Poole, F.~Sciortino, and H.E. Stanley.
\newblock {\em J. Chem. Phys.}, 107:7443, 1997.

\bibitem{vecchio}
H.~{Fraenkel-Conrat} and R.~C. {Williams}.
\newblock {\em Proc. Natl. Acad. Sci. U.S.A}, 41:690, 1955.

\bibitem{Zhang_04}
Z.~Zhang and S.~C. Glotzer.
\newblock {\em Nano Lett.}, {\bf 4}(8):1407--1413, 2004.

\bibitem{Workum_06}
Van Workum and J.~F. Douglas.
\newblock {\em Phys. Rev. E}, 73:031502, 2006.

\bibitem{zukosky}
C.F. Zukoski, G.~He, R.B.H. Tan, and P.J.A. Kenis.
\newblock {\em preprint}, 2007.

\bibitem{Poo92a}
P.~H. Poole, F.~Sciortino, U.~Essmann, and H.~E. Stanley.
\newblock {\em Nature}, 360:324--328,1992.

\bibitem{PNASstanley}
L.Xu, P.~Kumar, S.~V. Buldyrev, S.-H. Chen, P.~H. Poole, F.~Sciortino, and
  H.~E. Stanley.
\newblock {\em Proceedings Nat. Ac. Science}, 102:16558--16562, 2005.

\end{thebibliography}
\end{document}